# Performance Analysis of Linear-Equality-Constrained Least-Squares Estimation

Reza Arablouei and Kutluyıl Doğançay

*Abstract*—We analyze the performance of a linear-equality-constrained least-squares (CLS) algorithm and its relaxed version, called rCLS, that is obtained via the method of weighting. The rCLS algorithm solves an unconstrained least-squares problem that is augmented by incorporating a weighted form of the linear constraints. As a result, unlike the CLS algorithm, the rCLS algorithm is amenable to our approach to performance analysis presented here, which is akin to the energy-conservation-based methodology. Therefore, we initially inspect the convergence properties and evaluate the precision of estimation as well as satisfaction of the constraints for the rCLS algorithm in both mean and mean-square senses. Afterwards, we examine the performance of the CLS algorithm by evaluating the limiting performance of the rCLS algorithm as the relaxation parameter (weight) approaches infinity. Numerical examples verify the accuracy of the theoretical findings.

## I. INTRODUCTION

LINEARLY-CONSTRAINED adaptive filtering, whereby the adaptive filter coefficients are estimated subject to linear equality constraints, finds application in a wide range of areas such as spectral analysis, antenna array processing, spatial-temporal processing, blind multiuser detection, and linear-phase system identification. The constraints are generally deterministic and usually stem from prior knowledge of the underlying system, e.g., directions of arrival in array processing, spreading codes in blind multiuser detection, and linear phase in system identification. Moreover, in some applications, imposing particular linear equality constraints on the parameter estimates can provide specific advantages such as improving robustness of the estimates and obviating a training phase [1]-[3].

In [4], the first linearly-constrained adaptive filtering algorithm, called constrained least mean-square (CLMS), was proposed. It was initially conceived as an adaptive linearly-constrained minimum-variance filter for antenna array processing [2]. The CLMS algorithm is a stochastic-gradient-descent algorithm. It is simple and has a relatively low computational complexity [31]-[35]. However, it converges slowly, especially with correlated input. The constrained fast least-squares (CFLS) algorithm proposed in [5] converges faster than the CLMS algorithm but at the expense of increased computational complexity. A reduced-complexity but relaxed-constrains equivalent version of this algorithm has been proposed in [6], [7]. A recursive linearly-constrained total least-squares algorithm has been proposed in [8]. The constrained conjugate gradient, constrained affine projection, and constrained line-search algorithms, proposed in [9], [10], and [11], respectively, are also among other important linearly-constrained adaptive filtering algorithms.

An alternative way of realizing adaptive linear-equality-constrained estimation is to use the generalized sidelobe canceler (GSC) in conjunction with a conventional adaptive filtering algorithm [12]. This method requires multiplication of the input vector by a blocking matrix that is orthogonal to the constraint matrix, at each iteration. As a result, it is computationally more demanding than the direct-form linearly-constrained adaptive filtering algorithms, e.g., CLMS and CFLS, which have the constraints integrated into their update equations. Householder-transform constrained adaptive filtering [13] is an efficient implementation of the GSC that takes advantage of the efficiency of the Householder transformation to alleviate the computational complexity. It is shown in [14] that the GSC using the recursive least-squares algorithm is in fact mathematically equivalent to the CFLS algorithm.

Update equations of the linearly-constrained adaptive filtering algorithms are typically much more complicated than those of their unconstrained counterparts. Therefore, analyzing the performance of linearly-constrained adaptive filters is often very challenging. In [7], we applied the method of weighting [15] to convert the constrained least-squares (CLS) estimation problem to a new least-squares problem that is seemingly unconstrained while embedding the constraints in a relaxed manner. We achieve this by introducing a relaxation parameter (weight) and incorporating the constraints into the constraint-free form of the original least-squares problem. The solution of the resultant least-squares estimation problem is a relaxed CLS (rCLS) estimate that can be made arbitrarily close to the CLS estimate by increasing the weight. The CLS estimate is derived from the original CLS problem by applying the method of Lagrange multipliers. In [7], we also developed a low-complexity implementation of the rCLS algorithm employing the dichotomous coordinate-descent (DCD) iterations [16]-[18].

The rCLS algorithm in fact treats the equality constraints as extra measurements with certainty quantified by the weight. Such a treatment can be of practical significance, particularly when the parameters of the equality constraints are not precisely known [30]. The rCLS algorithm can be considered

R. Arablouei and K. Doğançay are with the School of Engineering and the Institute for Telecommunications Research, University of South Australia, Mawson Lakes SA 5095, Australia (email: arary003@mymail.unisa.edu.au, kutluyil.dogancay@unisa.edu.au).

as a generalized case of the CLS algorithm in the sense that it turns into the CLS algorithm when the weight tends to infinity. Moreover, the update equation of the rCLS algorithm is substantially more tractable than that of the CLS algorithm or any variant of it such as the CFLS algorithm, as far as performance analysis is concerned.

In this paper, we analyze the performance of the rCLS algorithm by adopting an approach motivated by the energy conservation arguments [19]-[21]. Then, we study the performance of the CLS algorithm by examining the analysis results of the rCLS algorithm when the value of the weight is sufficiently large to make the rCLS and CLS estimates identical. This is theoretically accomplished by computing the limits as the weight goes to infinity. In consequence, we gain valuable insights into the performance, i.e., convergence properties and estimation accuracy, of the CLS algorithm in an indirect way while any direct way appears to be formidable. The presented analysis also sheds light on the useful trade-offs offered by the rCLS algorithm in terms of convergence speed, estimation accuracy, and satisfaction of the constraints.

## II. Algorithms

Consider a linear system with an unobservable parameter vector $\mathbf{h} \in \mathbb{R}^{L \times 1}$ that at each time instant $n \in \mathbb{N}$ relates an input vector $\mathbf{x}_n \in \mathbb{R}^{L \times 1}$ to an output scalar $y_n \in \mathbb{R}$ through

$$y_n = \mathbf{x}_n^\top \mathbf{h} + v_n \quad (1)$$

where $v_n \in \mathbb{R}$ is noise and $L \in \mathbb{N}$ is the system order. A least-squares adaptive finite-impulse-response filter whose tap-coefficients are denoted by $\mathbf{w}_n \in \mathbb{R}^{L \times 1}$ is used to estimate $\mathbf{h}$ from the observable input-output data. Additionally, at every iteration, $\mathbf{w}_n$ is required to satisfy a set of $1 < K < L$ linear equality constraints such that

$$\mathbf{C}^\top \mathbf{w}_n = \mathbf{f} \quad (2)$$

where $\mathbf{C} \in \mathbb{R}^{L \times K}$ is the constraint matrix and $\mathbf{f} \in \mathbb{R}^{K \times 1}$ is the response vector.

Formally, we have

$$\mathbf{w}_n = \arg\min_{\mathbf{w}} \|\mathbf{y}_n - \mathbf{X}_n^\top \mathbf{w}\|^2 \text{ subject to } \mathbf{C}^\top \mathbf{w} = \mathbf{f} \quad (3)$$

where

$$\mathbf{X}_n = [\lambda^{(n-1)/2} \mathbf{x}_1, \lambda^{(n-2)/2} \mathbf{x}_2, \dots, \lambda^{1/2} \mathbf{x}_{n-1}, \mathbf{x}_n],$$

$$\mathbf{y}_n = [\lambda^{(n-1)/2} y_1, \lambda^{(n-2)/2} y_2, \dots, \lambda^{1/2} y_{n-1}, y_n]^\top,$$

$\|\cdot\|$ denotes the Euclidean norm, and $0 \ll \lambda < 1$ is a forgetting factor. Define the exponentially-weighted input covariance matrix as

$$\boldsymbol{\Phi}_n = \mathbf{X}_n \mathbf{X}_n^\top = \lambda \boldsymbol{\Phi}_{n-1} + \mathbf{x}_n \mathbf{x}_n^\top \quad (4)$$

and the exponentially-weighted input-output cross-covariance vector as

$$\mathbf{p}_n = \mathbf{X}_n \mathbf{y}_n = \lambda \mathbf{p}_{n-1} + y_n \mathbf{x}_n. \quad (5)$$

Solving the constrained minimization problem in (3) using the method of Lagrange multipliers results in [5]

$$\mathbf{w}_n = \boldsymbol{\Phi}_n^{-1} \mathbf{p}_n + \boldsymbol{\Phi}_n^{-1} \mathbf{C} (\mathbf{C}^\top \boldsymbol{\Phi}_n^{-1} \mathbf{C})^{-1} (\mathbf{f} - \mathbf{C}^\top \boldsymbol{\Phi}_n^{-1} \mathbf{p}_n). \quad (6)$$

We call (4)-(6) the constrained least-squares (CLS) algorithm. In [5], a low-complexity recursive implementation of this algorithm, termed the CFLS algorithm, is devised utilizing the Sherman-Morrison formula [22] and the stabilized fast transversal filter of [23].

Relaxing the constrained optimization problem of (3) using the method of weighting [15], yields

$$\mathbf{w}_n = \arg\min_{\mathbf{w}} \left\| \begin{bmatrix} \mathbf{y}_n \\ \mu^{1/2} \mathbf{f} \end{bmatrix} - \begin{bmatrix} \mathbf{X}_n^\top \\ \mu^{1/2} \mathbf{C}^\top \end{bmatrix} \mathbf{w} \right\|^2$$

where $\mu \gg 1$ is the relaxation parameter that we will refer to as the weight. As a result, $\mathbf{w}_n$ can alternatively be computed as

$$\mathbf{w}_n = (\boldsymbol{\Phi}_n + \mu \mathbf{C} \mathbf{C}^\top)^{-1} (\mathbf{p}_n + \mu \mathbf{C} \mathbf{f}). \quad (7)$$

We call (4), (5), and (7) the relaxed CLS (rCLS) algorithm.

Using the matrix identities [27], [28]

$$(\mathfrak{A} - \mathfrak{B} \mathfrak{D}^{-1} \mathfrak{C})^{-1} = \mathfrak{A}^{-1} + \mathfrak{A}^{-1} \mathfrak{B} (\mathfrak{D} - \mathfrak{C} \mathfrak{A}^{-1} \mathfrak{B})^{-1} \mathfrak{C} \mathfrak{A}^{-1} \quad (8)$$

and

$$(\mathfrak{A} + \mathfrak{B} \mathfrak{C})^{-1} \mathfrak{B} = \mathfrak{A}^{-1} \mathfrak{B} (\mathbf{I} + \mathfrak{C} \mathfrak{A}^{-1} \mathfrak{B})^{-1}$$

along with considering that

$$\lim_{\mu \to \infty} \mu^{-1} \mathbf{I} + \mathbf{C}^\top \boldsymbol{\Phi}_n^{-1} \mathbf{C} = \mathbf{C}^\top \boldsymbol{\Phi}_n^{-1} \mathbf{C},$$

we have

$$(\boldsymbol{\Phi}_n + \mu \mathbf{C} \mathbf{C}^\top)^{-1} \mathbf{p}_n = [\boldsymbol{\Phi}_n^{-1} - \boldsymbol{\Phi}_n^{-1} \mathbf{C} (\mu^{-1} \mathbf{I} + \mathbf{C}^\top \boldsymbol{\Phi}_n^{-1} \mathbf{C})^{-1} \mathbf{C}^T \boldsymbol{\Phi}_n^{-1}] \mathbf{p}_n \quad (9) = [\boldsymbol{\Phi}_n^{-1} - \boldsymbol{\Phi}_n^{-1} \mathbf{C} (\mathbf{C}^T \boldsymbol{\Phi}_n^{-1} \mathbf{C})^{-1} \mathbf{C}^T \boldsymbol{\Phi}_n^{-1}] \mathbf{p}_n$$

and

$$(\boldsymbol{\Phi}_n + \mu \mathbf{C} \mathbf{C}^\top)^{-1} \mu \mathbf{C} \mathbf{f} = \mu \boldsymbol{\Phi}_n^{-1} \mathbf{C} (\mathbf{I} + \mu \mathbf{C}^\top \boldsymbol{\Phi}_n^{-1} \mathbf{C})^{-1} \mathbf{f} = \boldsymbol{\Phi}_n^{-1} \mathbf{C} (\mathbf{C}^\top \boldsymbol{\Phi}_n^{-1} \mathbf{C})^{-1} \mathbf{f}. \quad (10)$$

Summation of (9) and (10) confirms that the rCLS estimate of (7) converges to the CLS estimate of (6) as $\mu$ approaches infinity.

The computational complexity of (7) is $\mathcal{O}(L^3)$. However, one can find $\mathbf{w}_n$ by solving the following system of linear equations utilizing the DCD algorithm:

$$(\boldsymbol{\Phi}_n + \mu \mathbf{C} \mathbf{C}^\top) \mathbf{w}_n = \mathbf{p}_n + \mu \mathbf{C} \mathbf{f}. \quad (11)$$

This reduced the computational complexity of the rCLS algorithm significantly, as detailed in [7]. In Table I, we present the total number of multiplications and additions per iteration required by the CFLS algorithm of [5], which is a fast version of the CLS algorithm, and the rCLS algorithm implemented using the DCD iterations, called the DCD-rCLS algorithm. We consider both cases of non-shift-structured and shift-structured input [7]. The DCD algorithm exercises maximum $N$ iterations and uses $M$ bits to represent each entry of the solution vector as a fixed-point word [16]-[18].

### III. Pre-analysis

In this section, we state the assumptions, define the performance measures, and derive the relevant recursive update equations to be used in our analysis.

#### A. Assumptions

We adopt the following assumptions, which are commonly used to facilitate the analytical studies [19], [24]:

*A1*: The input vector $\mathbf{x}_n$ is temporally independent with
$$E[\mathbf{x}_n] = \mathbf{0}, \ E[\mathbf{x}_n \mathbf{x}_n^\top] = \mathbf{R} \in \mathbb{R}^{L \times L}$$
where $\mathbf{0}$ denotes the $L \times 1$ zero vector and $\mathbf{R}$ is symmetric positive-definite.

*A2*: The noise $v_n$ is temporally independent with
$$E[v_n] = 0, \ E[v_n^2] = \eta \in \mathbb{R}_{\geq 0}.$$
In addition, it is independent of the input data.

*A3*: When $\lambda$ is close to 1, for a sufficiently large $n$, we can replace $\mathbf{\Phi}_n$ with its asymptotic expected value that, under *A1*, is calculated as
$$\lim_{n \to \infty} E[\mathbf{\Phi}_n] = (1 - \lambda)^{-1} \mathbf{R}.$$

The following corollary is deduced from (4)-(7) and *A1*-*A2*:

*C1*: The rCLS estimate of (7) at time instant $n-1$, i.e., $\mathbf{w}_{n-1}$, is independent of the input vector and noise at time instant $n$, i.e., $\mathbf{x}_n$ and $v_n$.

#### B. Recursive Update Equation: Deviation

At time instants $n-1$, (11) can be written as
$$(\mathbf{\Phi}_{n-1} + \mu \mathbf{C}\mathbf{C}^\top)\mathbf{w}_{n-1} = \mathbf{p}_{n-1} + \mu \mathbf{C}\mathbf{f}. \quad (12)$$

Multiplying (12) by $\lambda$ and substituting (4) and (5) into the resulting equation gives
$$(\mathbf{\Phi}_n - \mathbf{x}_n \mathbf{x}_n^\top + \lambda \mu \mathbf{C}\mathbf{C}^\top)\mathbf{w}_{n-1} = \mathbf{p}_n - y_n \mathbf{x}_n + \lambda \mu \mathbf{C}\mathbf{f}. \quad (13)$$

Subtracting (13) from (11) yields
$$(\mathbf{\Phi}_n + \mu \mathbf{C}\mathbf{C}^\top)\mathbf{w}_n = (\mathbf{\Phi}_n + \mu \mathbf{C}\mathbf{C}^\top)\mathbf{w}_{n-1} \\ - (\mathbf{x}_n \mathbf{x}_n^\top + \acute{\lambda} \mu \mathbf{C}\mathbf{C}^\top)\mathbf{w}_{n-1} + y_n \mathbf{x}_n + \acute{\lambda}\mu \mathbf{C}\mathbf{f} \quad (14)$$

where
$$\acute{\lambda} = 1 - \lambda.$$

Under *A1*-*A2*, the optimal constrained least-squares solution is given as [1]
$$\mathbf{g} = \mathbf{h} + \mathbf{R}^{-1}\mathbf{C}(\mathbf{C}^\top \mathbf{R}^{-1} \mathbf{C})^{-1}(\mathbf{f} - \mathbf{C}^\top \mathbf{h}). \quad (15)$$

Thus, we define the *deviation* vector as
$$\mathbf{d}_n = \mathbf{w}_n - \mathbf{g}.$$

Subtracting $(\mathbf{\Phi}_n + \mu \mathbf{C}\mathbf{C}^\top)\mathbf{g}$ from both sides of (14) together with using (1) gives
$$(\mathbf{\Phi}_n + \mu \mathbf{C}\mathbf{C}^\top)\mathbf{d}_n = (\mathbf{\Phi}_n + \mu \mathbf{C}\mathbf{C}^\top)\mathbf{d}_{n-1} \\ - (\mathbf{x}_n \mathbf{x}_n^\top + \acute{\lambda}\mu \mathbf{C}\mathbf{C}^\top)\mathbf{d}_{n-1} - (\mathbf{x}_n \mathbf{x}_n^\top + \acute{\lambda}\mu \mathbf{C}\mathbf{C}^\top)\mathbf{g} \\ + \mathbf{x}_n \mathbf{x}_n^\top \mathbf{h} + v_n \mathbf{x}_n + \acute{\lambda}\mu \mathbf{C}\mathbf{f}. \quad (16)$$

Using *A3* and considering that $\mathbf{C}^\top \mathbf{g} = \mathbf{f}$, (16) becomes
$$(\mathbf{R} + \acute{\lambda}\mu \mathbf{C}\mathbf{C}^\top)\mathbf{d}_n = (\mathbf{R} + \acute{\lambda}\mu \mathbf{C}\mathbf{C}^\top)\mathbf{d}_{n-1} \\ - \acute{\lambda}(\mathbf{x}_n \mathbf{x}_n^\top + \acute{\lambda}\mu \mathbf{C}\mathbf{C}^\top)\mathbf{d}_{n-1} \\ + \acute{\lambda}[\mathbf{x}_n \mathbf{x}_n^\top(\mathbf{h} - \mathbf{g}) + v_n \mathbf{x}_n]. \quad (17)$$

Defining
$$\mathbf{A} = (\mathbf{R} + \acute{\lambda}\mu \mathbf{C}\mathbf{C}^\top)^{-1} \quad (18)$$
and
$$\mathbf{e} = \mathbf{h} - \mathbf{g} \quad (19)$$
and multiplying (17) by $\mathbf{A}$ from the left, we arrive at
$$\mathbf{d}_n = [\mathbf{I} - \acute{\lambda}\mathbf{A}(\mathbf{x}_n \mathbf{x}_n^\top + \acute{\lambda}\mu \mathbf{C}\mathbf{C}^\top)]\mathbf{d}_{n-1} \\ + \acute{\lambda}\mathbf{A}(\mathbf{x}_n \mathbf{x}_n^\top \mathbf{e} + v_n \mathbf{x}_n) \quad (20)$$

where $\mathbf{I}$ is the $L \times L$ identity matrix.

#### C. Recursive Update Equation: Mismatch

To gauge the satisfaction of the constraints, we define the *mismatch* vector as
$$\mathbf{m}_n = \mathbf{C}^\top \mathbf{w}_n - \mathbf{f}.$$

This vector is not necessarily zero at all iterations of the rCLS algorithm. Therefore, it is useful to study its dynamics and dependence on different variables and parameters.

Substituting (1) into (14) and multiplying it by $\mathbf{C}^\top \mathbf{\Phi}_n^{-1}$ from the left gives
$$(\mathbf{I} + \mu \mathbf{C}^\top \mathbf{\Phi}_n^{-1}\mathbf{C})\mathbf{C}^\top \mathbf{w}_n = (\mathbf{I} + \lambda\mu \mathbf{C}^\top \mathbf{\Phi}_n^{-1}\mathbf{C})\mathbf{C}^\top \mathbf{w}_{n-1} \\ - \mathbf{C}^\top \mathbf{\Phi}_n^{-1}\mathbf{x}_n \mathbf{x}_n^\top \mathbf{w}_{n-1} + \mathbf{C}^\top \mathbf{\Phi}_n^{-1}\mathbf{x}_n \mathbf{x}_n^\top \mathbf{h} \\ + \mathbf{C}^\top \mathbf{\Phi}_n^{-1}v_n \mathbf{x}_n + \acute{\lambda}\mu \mathbf{C}^\top \mathbf{\Phi}_n^{-1}\mathbf{C}\mathbf{f}. \quad (21)$$

Subtracting $(\mathbf{I} + \mu \mathbf{C}^\top \mathbf{\Phi}_n^{-1}\mathbf{C})\mathbf{f}$ from both sides of (21) results in
$$(\mathbf{I} + \mu \mathbf{C}^\top \mathbf{\Phi}_n^{-1}\mathbf{C})\mathbf{m}_n = (\mathbf{I} + \lambda\mu \mathbf{C}^\top \mathbf{\Phi}_n^{-1}\mathbf{C})\mathbf{m}_{n-1} \\ - \mathbf{C}^\top \mathbf{\Phi}_n^{-1}\mathbf{x}_n \mathbf{x}_n^\top \mathbf{w}_{n-1} + \mathbf{C}^\top \mathbf{\Phi}_n^{-1}\mathbf{x}_n^\top \mathbf{x}_n \mathbf{h} \\ + v_n \mathbf{C}^\top \mathbf{\Phi}_n^{-1}\mathbf{x}_n. \quad (22)$$

Using *A3* and approximating $\mathbf{x}_n \mathbf{x}_n^\top$ with its expectation, i.e., $\mathbf{R}$, (22) can be approximately written as

$$(\mathbf{I} + \acute{\lambda}\mu\mathbf{C}^\top\mathbf{R}^{-1}\mathbf{C})\mathbf{m}_n = \lambda(\mathbf{I} + \acute{\lambda}\mu\mathbf{C}^\top\mathbf{R}^{-1}\mathbf{C})\mathbf{m}_{n-1} \\ + \acute{\lambda}(\mathbf{C}^\top\mathbf{h} - \mathbf{f}) + \acute{\lambda}v_n\mathbf{C}^\top\mathbf{R}^{-1}\mathbf{x}_n. \quad (23)$$

Defining

$$\mathbf{B} = (\mathbf{I} + \acute{\lambda}\mu\mathbf{C}^\top\mathbf{R}^{-1}\mathbf{C})^{-1} \quad (24)$$

and

$$\mathbf{r} = \mathbf{C}^\top\mathbf{h} - \mathbf{f}$$

and multiplying (23) by $\mathbf{B}$ from the left, we get

$$\mathbf{m}_n = \lambda\mathbf{m}_{n-1} + \acute{\lambda}\mathbf{B}\mathbf{r} + \acute{\lambda}v_n\mathbf{B}\mathbf{C}^\top\mathbf{R}^{-1}\mathbf{x}_n. \quad (25)$$

## IV. Performance of rCLS

In this section, we analyze the mean and mean-square performance of the rCLS algorithm.

### A. Mean Deviation

Taking the expectation on both sides of (20) with *A2* and *C1* in mind gives

$$E[\mathbf{d}_n] = \lambda E[\mathbf{d}_{n-1}] + \acute{\lambda}\mathbf{A}\mathbf{R}\mathbf{e}. \quad (26)$$

As $\lambda < 1$, (26) converges and the asymptotic bias of the rCLS algorithm is given by

$$\lim_{n\to\infty} E[\mathbf{d}_n] = \mathbf{A}\mathbf{R}\mathbf{e}. \quad (27)$$

### B. Mean-Square Stability

Taking the expectation of the squared Euclidean norm on both sides of (20) with *C1* in mind gives

$$E[\|\mathbf{d}_n\|^2] = E[\mathbf{d}_{n-1}^\top\mathbf{S}\mathbf{d}_{n-1}] \\ + \acute{\lambda}^2 E[(\mathbf{e}^\top\mathbf{x}_n\mathbf{x}_n^\top + v_n\mathbf{x}_n^\top)\mathbf{A}^2(\mathbf{x}_n\mathbf{x}_n^\top\mathbf{e} + v_n\mathbf{x}_n)] \\ + 2\acute{\lambda}E[(\mathbf{e}^\top\mathbf{x}_n\mathbf{x}_n^\top + v_n\mathbf{x}_n^\top)\mathbf{A} \\ \times \{\mathbf{I} - \acute{\lambda}\mathbf{A}(\mathbf{x}_n\mathbf{x}_n^\top + \acute{\lambda}\mu\mathbf{C}\mathbf{C}^\top)\}]E[\mathbf{d}_{n-1}] \quad (28)$$

where

$$\mathbf{S} = E[\{\mathbf{I} - \acute{\lambda}(\mathbf{x}_n\mathbf{x}_n^\top + \acute{\lambda}\mu\mathbf{C}\mathbf{C}^\top)\mathbf{A}\}\{\mathbf{I} - \acute{\lambda}\mathbf{A}(\mathbf{x}_n\mathbf{x}_n^\top + \acute{\lambda}\mu\mathbf{C}\mathbf{C}^\top)\}].$$

Using the Isserlis' theorem [25] and *A1*, we have

$$E[\mathbf{x}_n\mathbf{x}_n^\top\mathbf{A}^2\mathbf{x}_n\mathbf{x}_n^\top] = E[\mathbf{x}_n^\top\mathbf{A}^2\mathbf{x}_n]E[\mathbf{x}_n\mathbf{x}_n^\top] \\ + 2E[\mathbf{x}_n\mathbf{x}_n^\top]\mathbf{A}^2 E[\mathbf{x}_n\mathbf{x}_n^\top] \quad (29) \\ = \mathrm{tr}\{\mathbf{A}^2\mathbf{R}\}\mathbf{R} + 2\mathbf{R}\mathbf{A}^2\mathbf{R}$$

hence

$$\mathbf{S} = (2\lambda - 1)\mathbf{I} \\ + \acute{\lambda}^2(\mathrm{tr}\{\mathbf{A}^2\mathbf{R}\}\mathbf{R} + \acute{\lambda}^2\mu^2\mathbf{C}\mathbf{C}^\top\mathbf{A}^2\mathbf{C}\mathbf{C}^\top + \mathbf{R}\mathbf{A} + \mathbf{A}\mathbf{R}). \quad (30)$$

The variance relation (28) is stable if the spectral radius of $\mathbf{S}$ is less than one, i.e.,

$$\rho\{\mathbf{S}\} = 2\lambda - 1 \\ + \acute{\lambda}^2\rho\{\mathrm{tr}\{\mathbf{A}^2\mathbf{R}\}\mathbf{R} + \acute{\lambda}^2\mu^2\mathbf{C}\mathbf{C}^\top\mathbf{A}^2\mathbf{C}\mathbf{C}^\top + \mathbf{R}\mathbf{A} + \mathbf{A}\mathbf{R}\} \\ < 1.$$

Therefore, considering the sub-additive inequality of the spectral radius for any two multiplication-commutative matrices [26], i.e.,

$$\rho\{\mathfrak{A} + \mathfrak{B}\} \leq \rho\{\mathfrak{A}\} + \rho\{\mathfrak{B}\},$$

where $\mathfrak{A}\mathfrak{B} = \mathfrak{B}\mathfrak{A}$, the mean-square stability of the rCLS algorithm requires

$$\acute{\lambda}(\mathrm{tr}\{\mathbf{A}^2\mathbf{R}\}\rho\{\mathbf{R}\} + \acute{\lambda}^2\mu^2\rho\{\mathbf{C}\mathbf{C}^\top\mathbf{A}^2\mathbf{C}\mathbf{C}^\top\} + 2\rho\{\mathbf{A}\mathbf{R}\}) < 2. \quad (31)$$

### C. Steady-State Mean-Square Deviation

Using the Isserlis' theorem and *A1*, we have

$$E[\mathbf{x}_n\mathbf{x}_n^\top\mathbf{e}\mathbf{e}^\top\mathbf{x}_n\mathbf{x}_n^\top] = E[(\mathbf{e}^\top\mathbf{x}_n)(\mathbf{x}_n^\top\mathbf{e})\mathbf{x}_n\mathbf{x}_n^\top] \\ = E[(\mathbf{e}^\top\mathbf{x}_n)(\mathbf{x}_n^\top\mathbf{e})]E[\mathbf{x}_n\mathbf{x}_n^\top] \\ + 2E[(\mathbf{x}_n^\top\mathbf{e})\mathbf{x}_n]E[(\mathbf{e}^\top\mathbf{x}_n)\mathbf{x}_n^\top] \quad (32) \\ = (\mathbf{e}^\top\mathbf{R}\mathbf{e})\mathbf{R} + 2\mathbf{R}\mathbf{e}\mathbf{e}^\top\mathbf{R}.$$

Using (29) and (32) in (28) and ignoring the second additive term on the right-hand side of (30) due to the small value of $\acute{\lambda}^2$ as $\lambda$ is set close to one, we obtain

$$E[\|\mathbf{d}_n\|^2] = (2\lambda - 1)E[\|\mathbf{d}_{n-1}\|^2] \\ + \acute{\lambda}^2\mathrm{tr}\{\mathbf{A}^2[(\mathbf{e}^\top\mathbf{R}\mathbf{e} + \eta)\mathbf{R} + 2\mathbf{R}\mathbf{e}\mathbf{e}^\top\mathbf{R}]\} \quad (33) \\ + 2\acute{\lambda}\mathbf{e}^\top\mathbf{R}[\lambda\mathbf{A} - \acute{\lambda}(\mathrm{tr}\{\mathbf{A}^2\mathbf{R}\}\mathbf{I} + \mathbf{A}^2\mathbf{R})]E[\mathbf{d}_{n-1}].$$

Since $\lambda < 1$, we have $(2\lambda - 1) < 1$. Thus, (33) converges and, using (27), the steady-state mean-square deviation (MSD) of the rCLS algorithm is given by

$$\lim_{n\to\infty} E[\|\mathbf{d}_n\|^2] = \frac{\acute{\lambda}}{2}\mathrm{tr}\{\mathbf{A}^2[(\mathbf{e}^\top\mathbf{R}\mathbf{e} + \eta)\mathbf{R} + 2\mathbf{R}\mathbf{e}\mathbf{e}^\top\mathbf{R}]\} \\ + \mathbf{e}^\top\mathbf{R}[\lambda\mathbf{A} - \acute{\lambda}(\mathrm{tr}\{\mathbf{A}^2\mathbf{R}\}\mathbf{I} + \mathbf{A}^2\mathbf{R})]\mathbf{A}\mathbf{R}\mathbf{e}. \quad (34)$$

### D. Mean and Mean-Square Mismatch

Taking the expectation on both sides of (25) while considering *A2* gives

$$E[\mathbf{m}_n] = \lambda E[\mathbf{m}_{n-1}] + \acute{\lambda}\mathbf{B}\mathbf{r}.$$

Therefore, we have

$$\lim_{n\to\infty} E[\mathbf{m}_n] = \mathbf{B}\mathbf{r}. \quad (35)$$

In view of *A2* and *C1*, taking the expectation of the squared Euclidean norm on both sides of (25) results in

$$E[\|\mathbf{m}_n\|^2] = \lambda^2 E[\|\mathbf{m}_{n-1}\|^2] + \acute{\lambda}^2\mathbf{r}^\top\mathbf{B}^2\mathbf{r} \\ + \acute{\lambda}^2 E[v_n^2\mathbf{x}_n^\top\mathbf{R}^{-1}\mathbf{C}\mathbf{B}^2\mathbf{C}^\top\mathbf{R}^{-1}\mathbf{x}_n] + 2\lambda\acute{\lambda}\mathbf{r}^\top\mathbf{B}E[\mathbf{m}_{n-1}].$$

Consequently, considering (35), the steady-state mean-square mismatch (MSM) of the rCLS algorithm is given by

$$\lim_{n\to\infty} E[\|\mathbf{m}_n\|^2] = \left(\frac{1-\lambda}{1+\lambda}\right)\eta\,\mathrm{tr}\{\mathbf{R}^{-1}\mathbf{CB}^2\mathbf{C}^\top\} + \mathbf{r}^\top\mathbf{B}^2\mathbf{r}$$
$$= \mathrm{tr}\left\{\mathbf{B}^2\left[\left(\frac{1-\lambda}{1+\lambda}\right)\eta\mathbf{C}^\top\mathbf{R}^{-1}\mathbf{C} + \mathbf{rr}^\top\right]\right\}. \quad (36)$$

## V. Performance of CLS

In this section, we study the mean and mean-square performance of the CLS algorithm and calculate its performance metrics by letting the weight $\mu$ go to infinity in the performance analysis results derived for the rCLS algorithm in the previous section.

### A. Mean Deviation

Applying (8) to (18) gives

$$\mathbf{A} = \mathbf{R}^{-1} - \mathbf{R}^{-1}\mathbf{C}\big(\acute{\lambda}^{-1}\mu^{-1}\mathbf{I} + \mathbf{C}^\top\mathbf{R}^{-1}\mathbf{C}\big)^{-1}\mathbf{C}^\top\mathbf{R}^{-1}.$$

Subsequently, we have

$$\lim_{\mu\to\infty} \mathbf{A} = \mathbf{R}^{-1} - \mathbf{R}^{-1}\mathbf{C}(\mathbf{C}^\top\mathbf{R}^{-1}\mathbf{C})^{-1}\mathbf{C}^\top\mathbf{R}^{-1}. \quad (37)$$

Thus, (27), (15), (19), and (37) lead to

$$\lim_{n,\mu\to\infty} E[\mathbf{d}_n] = [\mathbf{I} - \mathbf{R}^{-1}\mathbf{C}(\mathbf{C}^\top\mathbf{R}^{-1}\mathbf{C})^{-1}\mathbf{C}^\top]$$
$$\times \mathbf{R}^{-1}\mathbf{C}(\mathbf{C}^\top\mathbf{R}^{-1}\mathbf{C})^{-1}(\mathbf{C}^\top\mathbf{h} - \mathbf{f}) \quad (38)$$
$$= \mathbf{0}.$$

This indicates that the CLS algorithm is asymptotically unbiased.

### B. Mean-Square Deviation

Using (37), we have

$$\lim_{\mu\to\infty} \mathbf{AR} = \mathbf{I} - \mathbf{R}^{-1}\mathbf{C}(\mathbf{C}^\top\mathbf{R}^{-1}\mathbf{C})^{-1}\mathbf{C}^\top \quad (39)$$

and

$$\lim_{\mu\to\infty} \mathbf{A}^2\mathbf{R} = \mathbf{R}^{-1} - \mathbf{R}^{-1}\mathbf{C}(\mathbf{C}^\top\mathbf{R}^{-1}\mathbf{C})^{-1}\mathbf{C}^\top\mathbf{R}^{-1}. \quad (40)$$

Using the identities [27], [28]

$$\mathfrak{BC}(\mathfrak{A} + \mathfrak{BC})^{-1} = \mathfrak{B}(\mathbf{I} + \mathfrak{CA}^{-1}\mathfrak{B})^{-1}\mathfrak{CA}^{-1}$$

and

$$(\mathfrak{A} + \mathfrak{BC})^{-1}\mathfrak{BC} = \mathfrak{A}^{-1}\mathfrak{B}(\mathbf{I} + \mathfrak{CA}^{-1}\mathfrak{B})^{-1}\mathfrak{C},$$

we also have

$$\acute{\lambda}\mu\mathbf{CC}^\top\mathbf{A} = \acute{\lambda}\mu\mathbf{C}\big(\mathbf{I} + \acute{\lambda}\mu\mathbf{C}^\top\mathbf{R}^{-1}\mathbf{C}\big)^{-1}\mathbf{C}^\top\mathbf{R}^{-1}$$

and

$$\acute{\lambda}\mu\mathbf{ACC}^\top = \acute{\lambda}\mu\mathbf{R}^{-1}\mathbf{C}\big(\mathbf{I} + \acute{\lambda}\mu\mathbf{C}^\top\mathbf{R}^{-1}\mathbf{C}\big)^{-1}\mathbf{C}^\top$$

hence

$$\lim_{\mu\to\infty} \acute{\lambda}\mu\mathbf{CC}^\top\mathbf{A} = \mathbf{C}(\mathbf{C}^\top\mathbf{R}^{-1}\mathbf{C})^{-1}\mathbf{C}^\top\mathbf{R}^{-1} \quad (41)$$

and

$$\lim_{\mu\to\infty} \acute{\lambda}\mu\mathbf{ACC}^\top = \mathbf{R}^{-1}\mathbf{C}(\mathbf{C}^\top\mathbf{R}^{-1}\mathbf{C})^{-1}\mathbf{C}^\top. \quad (42)$$

Substituting (39)–(42) in (31) gives

$$\acute{\lambda}(\mathrm{tr}\{(\mathbf{I} - \mathbf{G})\mathbf{R}^{-1}\}\rho\{\mathbf{R}\} + \rho\{\mathbf{G}^\top\mathbf{G}\} + 2\rho\{\mathbf{I} - \mathbf{G}\}) < 2 \quad (43)$$

where

$$\mathbf{G} = \mathbf{R}^{-1}\mathbf{C}(\mathbf{C}^\top\mathbf{R}^{-1}\mathbf{C})^{-1}\mathbf{C}^\top.$$

The matrix $\mathbf{G}$ is idempotent as $\mathbf{G}^2 = \mathbf{G}$. Therefore, $\mathbf{I} - \mathbf{G}$ is also idempotent and has a unit spectral radius [29]. Consequently, (43) can be written as

$$\frac{\mathrm{tr}\{(\mathbf{I} - \mathbf{G})\mathbf{R}^{-1}\}\rho\{\mathbf{R}\} + \rho\{\mathbf{G}^\top\mathbf{G}\}}{\mathrm{tr}\{(\mathbf{I} - \mathbf{G})\mathbf{R}^{-1}\}\rho\{\mathbf{R}\} + \rho\{\mathbf{G}^\top\mathbf{G}\} + 2} < \lambda < 1,$$

which presents a lower bound on $\lambda$ for ensuring mean-square stability of the CLS algorithm.

Substituting (38) and (37) into (34) yields the steady-state MSD of the CLS algorithm as

$$\lim_{n,\mu\to\infty} E[\|\mathbf{d}_n\|^2]$$
$$= \frac{\acute{\lambda}}{2}\mathrm{tr}\{[(\mathbf{I} - \mathbf{G})\mathbf{R}^{-1}]^2[(\mathbf{e}^\top\mathbf{Re} + \eta)\mathbf{R} + 2\mathbf{Ree}^\top\mathbf{R}]\}. \quad (44)$$

### C. Mean and Mean-Square Mismatch

Using (8) in (24) gives

$$\mathbf{B} = \acute{\lambda}^{-1}\mu^{-1}(\mathbf{C}^\top\mathbf{R}^{-1}\mathbf{C})^{-1}$$
$$\times \left\{\mathbf{I} - \acute{\lambda}^{-1}\mu^{-1}\big[\mathbf{I} + \acute{\lambda}^{-1}\mu^{-1}(\mathbf{C}^\top\mathbf{R}^{-1}\mathbf{C})^{-1}\big]^{-1}(\mathbf{C}^\top\mathbf{R}^{-1}\mathbf{C})^{-1}\right\}.$$

Moreover, we have

$$\lim_{\mu\to\infty}\big[\mathbf{I} + \acute{\lambda}^{-1}\mu^{-1}(\mathbf{C}^\top\mathbf{R}^{-1}\mathbf{C})^{-1}\big]$$
$$= \lim_{\mu\to\infty}\big[\mathbf{I} - \acute{\lambda}^{-1}\mu^{-1}(\mathbf{C}^\top\mathbf{R}^{-1}\mathbf{C})^{-1}\big]$$
$$= \mathbf{I}$$

and

$$\lim_{\mu\to\infty} \acute{\lambda}^{-1}\mu^{-1}(\mathbf{C}^\top\mathbf{R}^{-1}\mathbf{C})^{-1} = \mathbf{O}$$

where $\mathbf{O}$ denotes the $L \times L$ zero matrix. Therefore, it holds that

$$\lim_{\mu\to\infty} \mathbf{B} = \mathbf{O}. \quad (45)$$

Using (45) with (35) and (36), we get

$$\lim_{n,\mu\to\infty} E[\mathbf{m}_n] = \mathbf{0}$$

and

$$\lim_{n,\mu\to\infty} E[\|\mathbf{m}_n\|^2] = 0.$$

This means that at the steady state, unlike the rCLS algorithm, the CLS algorithm strictly fulfills the constraints in both mean and mean-square senses.

## VI. SIMULATIONS

We consider a problem of linearly-constrained system identification where the system order is either $L = 7$ or $31$ and the number of linear equality constraints is $K = (L-1)/2$. We choose the parameters, $\mathbf{h}$, $\mathbf{C}$, $\mathbf{f}$, and $\mathbf{R}$, arbitrarily with the condition that $\mathbf{h}$ has unit energy, $\mathbf{C}$ is full-rank, and $\mathbf{R}$ is symmetric positive-definite with $\mathrm{tr}\{\mathbf{R}\} = L$[1]. The input vector is drawn from a zero-mean multivariate Gaussian distribution in the scenario of $L = 7$ and from a zero-mean multivariate uniform distribution in the scenario of $L = 31$. The noise is zero-mean Gaussian and independent of the input vector. We obtain the experimental results by calculating ensemble averages over $10^4$ independent trials and compute the steady-state quantities by averaging over $10^3$ steady-state values.

In Figs. 1-4, we plot the steady-state MSD and MSM of the rCLS algorithm, i.e., (34) and (36), respectively, versus $\mu$ for different values of $\lambda$ when $\eta = 0.1$. In Figs. 5 and 6, we plot the steady-state MSD of the CLS algorithm, i.e., (44), as a function of $\eta$ for different values of $\lambda$. Figs. 1-6 show both theoretical and experimental results for the considered scenarios and indicate a good match between theory and experiment verifying the analytical performance results established in this paper. Expectedly, the steady-state MSD levels to which the rCLS algorithm converges in Figs. 1 and 3 as $\eta$ increases perfectly match the steady-state MSDs for the CLS algorithm in Figs. 5 and 6, respectively, for $\eta = 0.1$ ($-10$ dB).

## VII. CONCLUSION

We have presented the performance analysis of a relaxed linearly-constrained least-squares (rCLS) algorithm that is formulated utilizing the method of weighting. This algorithm is mathematically equivalent to the constrained least-squares (CLS) algorithm, derived from the method of Lagrange multipliers, when the relaxation parameter (weight) approaches infinity. Therefore, to project the analytical results obtained for the rCLS algorithm to the CLS algorithm, we took the limits as the weight goes to infinity. This in fact enabled us to ultimately accomplish a rigorous performance analysis for the CLS algorithm, which is otherwise intractable to direct analytical approaches due to its complicated update equation. We considered two performance measures, namely, deviation and mismatch vectors. The former is the difference between the current estimate and the optimal CLS solution and the latter represents the error in satisfaction of the constraints by the current estimate. We studied mean and mean-square behavior of both performance measures for the rCLS and CLS algorithms and calculated their steady-state values. Simulations demonstrated an excellent match between the theoretical predictions and the experimental results.

---

[1] The explicit values of $\mathbf{h}$, $\mathbf{C}$, $\mathbf{f}$, and $\mathbf{R}$ along with the MATLAB source code to regenerate the simulation results have been provided as the supplementary material.

TABLE I
COMPUTATIONAL COMPLEXITY OF THE CFLS AND DCD-rCLS ALGORITHMS IN TERMS OF THE NUMBER OF REQUIRED MULTIPLICATIONS AND ADDITIONS PER ITERATION.

| | $\times$ | $+$ |
|---|---|---|
| | non-shift-structured input | |
| CFLS | $4L^2 + (3K^2 + 5K + 4)L + K^2 + 2K$ | $3L^2 + (3K^2 + 4K + 1)L - K^2 + K$ |
| DCD-rCLS | $\frac{1}{2}L^2 + \left(2K + \frac{5}{2}\right)L$ | $\frac{3}{2}L^2 + \left(2K + 2N + \frac{11}{2}\right)L + N + M$ |
| | shift-structured input | |
| CFLS | $(3K^2 + 5K + 9)L + K^2 + 2K + 16$ | $(3K^2 + 4K + 8)L - K^2 + K - 1$ |
| DCD-rCLS | $(2K + 3)L$ | $\frac{1}{2}L^2 + \left(2K + 2N + \frac{13}{2}\right)L + N + M$ |

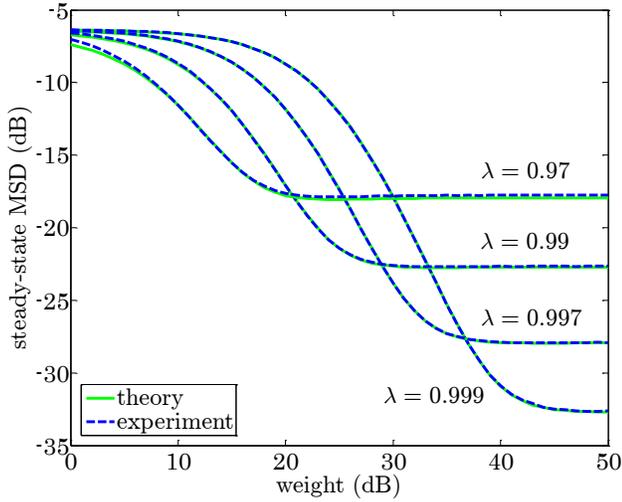

Fig. 1. Steady-state MSD of the rCLS algorithm versus $\mu$ for different values of $\lambda$ when $\eta = 0.1$ and $L = 7$.

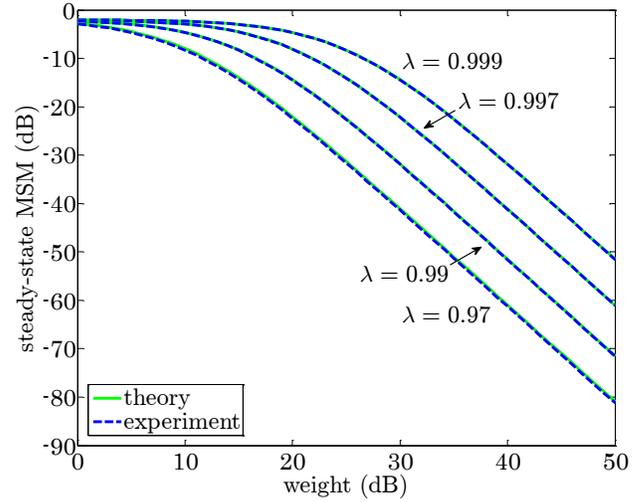

Fig. 2. Steady-state MSM of the rCLS algorithm versus $\mu$ for different values of $\lambda$ when $\eta = 0.1$ and $L = 7$.

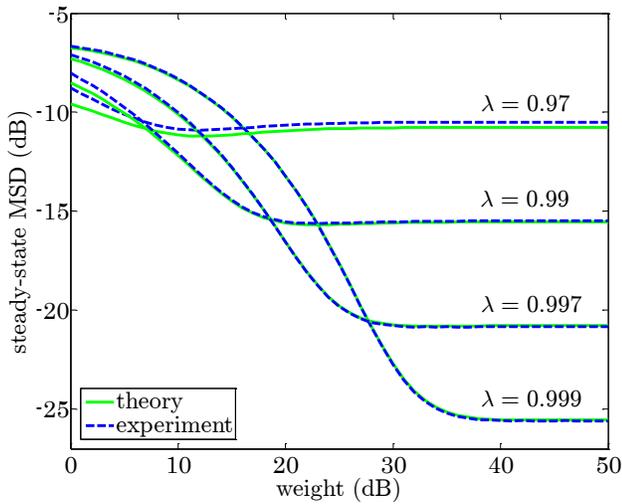

Fig. 3. Steady-state MSD of the rCLS algorithm versus $\mu$ for different values of $\lambda$ when $\eta = 0.1$ and $L = 31$.

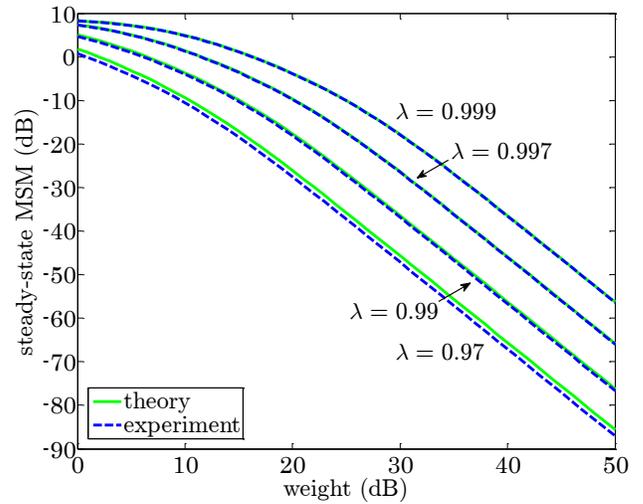

Fig. 4. Steady-state MSM of the rCLS algorithm versus $\mu$ for different values of $\lambda$ when $\eta = 0.1$ and $L = 31$.

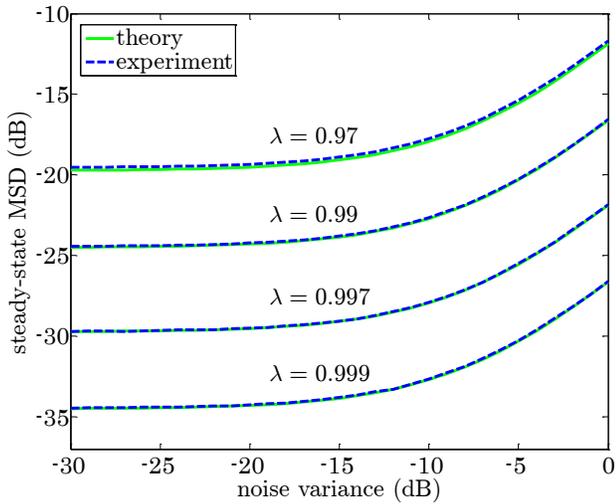

Fig. 5. Steady-state MSD of the CLS algorithm versus $\eta$ when $L = 7$.

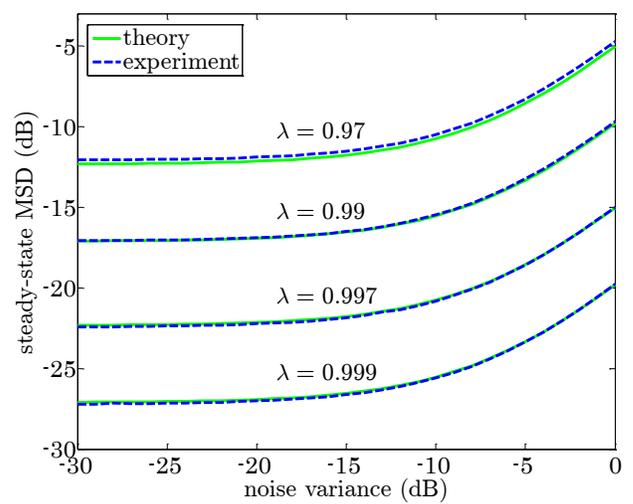

Fig. 6. Steady-state MSD of the CLS algorithm versus $\eta$ when $L = 31$.